\documentclass[pra,aps,showpas,floatfix,preprint]{revtex4}
\usepackage{graphicx}
\usepackage{bm} \usepackage{amssymb}

\begin{document}
\title{Bifurcations and bistability in cavity assisted photoassociation of Bose-Einstein condensed molecules}
\author{Markku~J{\"a}{\"a}skel{\"a}inen\footnote{Corresponding author.
Email address: mrq@phy.stevens.edu}, Jaeyoon Jeong, and Christopher~P.~Search}
\affiliation{Department of Physics and Engineering Physics,
Stevens Institute of Technology, Castle point on the Hudson, Hoboken, NJ 07030, USA}
\date{\today}
\begin{abstract}
We study the photo-association of Bose-Einstein condensed atoms into molecules using an optical cavity field.
The driven cavity field introduces a new dynamical degree of freedom into the photoassociation process, whose
role in determining the stationary behavior has not previously been considered. The semiclassical stationary solutions for the atom and
molecules as well as the intracavity field are found and their stability and scaling properties are determined in terms of
experimentally controllable parameters including driving amplitude of the cavity and the nonlinear interactions between
atoms and molecules. For weak cavity driving, we find a bifurcation in the atom and molecule
number occurs that signals a transition from a stable steady state to nonlinear Rabi oscillations.
For a strongly driven cavity, there exists bistability in the atom and molecule number.
\end{abstract}
\pacs{}
\maketitle

\section{Introduction}

Molecules can not be directly cooled using the laser cooling
techniques that led to Bose-Einstein condensation of alkali atoms
because of the complex rotational-vibrational spectrum of the
molecules. As a result, two-color Raman photoassociation and
Feshbach resonances have emerged as tools to create
translationally cold molecules starting from ultra-cold atomic
gases. The conversion of a macroscopic number of quantum
degenerate atoms into molecular dimers starting from either a
Bose-Einstein condensate (BEC) \cite{mol-1,durr,xu-2003,herbig} or a
Fermi gas \cite{regal-2003,strecker-2003,jochim-2003,cubizolles-2003} has
been observed by several experimental groups using Feshbach
resonances. This work culminated in the formation of a molecular
Bose-Einstein condensate (MBEC) \cite{mol-BEC-K,mol-BEC-Li}.

Although Feshbach resonances have been the most successful tool
for creating quantum degenerate gases of molecules, experiments
have demonstrated that two-color Raman photoassociation can also
be used to create molecules in the electronic ground state
\cite{wynar,julienne1,rom,winkler}. Two-color photoassociation has
the added benefit that the frequency difference between the two
optical fields can be used to select a particular
rotational-vibrational state \cite{tsai,jones}. This gives
photoassociation a potential advantage over Feshbach resonances
since photoassociation can be used to prepare molecules in their
rotational-vibrational ground state. While the molecules created
via a Feshbach resonance are translationally very cold, they are
vibrationally very hot and can decay to lower lying vibrational
states via exoergic inelastic collisions with atoms or other
molecules. For an atomic BEC, the molecular two-body decay rates
are of the order $10^{-11}-10^{-10}cm^3/s$, which gives a lifetime
of $100\mu s$ for typical atomic densities
\cite{xu-2003,yurovsky,mukaiyama}.

From the perspective of atom optics, the conversion of atoms into
molecules via a Feshbach resonance or photoassociation is
the matter-wave analog of second harmonic generation of
photons in a nonlinear crystal with a $\chi^{(2)}$ susceptibility,
which has been used to create entangled photon states. Recent
experiments have shown the phase coherent and momentum conserving
nature of matter-wave second harmonic generation
\cite{molecule-optics}. However, implicit in these analogies to
nonlinear optics is that the coupling strength between atoms and
molecules is not itself a quantum field.

Here we address the issue of two-photon Raman photoassociation
of an atomic BEC inside of an optical cavity. In this case one of the optical
fields used to induce the atom-molecule conversion is a quantized
mode of a driven optical resonator while the other field is
a laser with sufficient intensity to be treated as a 'classical'
undepleted pump. This is reminiscent of early work done by Moore and
Meystre \cite{moore} studying the coupling of a zero momentum BEC
atoms to atoms with momentum $\pm K$ due to a Raman transition
involving a quantized ring cavity mode and classical pump. Unlike their system, our
model involves the interaction of four particles: two atoms are
'destroyed' and a molecule and cavity photon are 'created' and
vice versa. Consequently, the atom-molecule-cavity interaction is
analogous to $\chi^{(3)}$ susceptibility in nonlinear optics, which is known to give
rise to four-wave mixing. Coherent photoassociation inside of cavity therefore offers the prospect of novel
nonlinear dynamics between the atomic, molecular, and cavity fields
as well as the possibility to entangle individual photons with
completely different chemical compounds- atoms and molecular
dimers.

In fact, only a few papers have previously considered photoassociation
inside of a cavity \cite{olsen1,olsen2}. However, unlike our model, theirs
was based on single photon photoassociation, which is impractical for observing coherent
atom-molecule dynamics because the molecules created are in
electronic excited states and can rapidly decay due to
spontaneous emission. The authors of Ref. \cite{olsen1,olsen2} employed a positive-P
distribution originally developed by Gardiner and Drummond
\cite{QuantumNoise} to analyze the quantum mechanicial 'phase space' dynamics of the
three coupled bosonic fields. Since their interaction
Hamiltonian required the absorption of a photon to create a
molecule, the equation of motion for the P-distribution involved
third order derivatives. Such derivatives meant that the resulting
equations of motion could not be identified as a Fokker-Planck
equation, which can be readily solved using standard techniques \cite{QuantumNoise}. As we will show here, Raman
photoassociation in which a cavity photon is created along with
the molecule leads to an equation of motion for the P-distribution
involving only first and second derivatives that is in the form of
a Fokker-Planck equation. Consequently, intracavity Raman
photoassociation leads not only to more stable molecules but also
equations of motion for quasi-phase space distributions that are more
easily solved using known techniques.

Here, we analyze the semiclassical behavior for the atomic, molecular, and cavity mean fields.
Our goal is to explore the stationary solutions of the
coupled nonlinear mean field equations  as well as the stability conditions for these solutions in terms of experimentally controllable parameters.
When the cavity is only weakly driven by an external source, the stationary intracavity field
remains close to zero while the molecules exhibit a saddle point bifurcation in the molecule number (or equivalently, atoms due to number
conservation). Below a critical driving amplitude, there exists both unstable and stable stationary solutions while above the critical
point there are no stationary solutions but instead only nonlinear Rabi oscillations. For strong cavity
driving, the system exhibits bistability in the number of molecules.
The paper is organized as follows: In section II, we present our
model for cavity assisted photoassociation and derive the equation of motion for the Positive-P
representation. From the resulting Fokker-Planck equation we obtain equations of motions
for the expectation values of the fields. In section
III, we analyze the solutions of these mean-field equations. In section IV, we discuss the implications of our results.

\section{Model}

\begin{figure}[ht]
\begin{center}
\includegraphics[width=8.5cm]{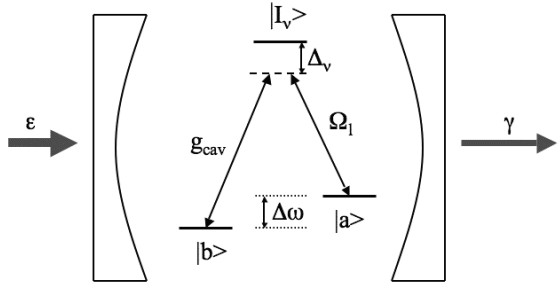} \caption{\label{Fig1}
Schematic diagram showing the system under consideration. The kets $|a\rangle$, $|b\rangle$, and $|I_\nu\rangle$ denote pairs of atoms, electronic ground state molecules, and electronically excited molecules, respectively. $\varepsilon$ is the rate at which the cavity is coherently driven by an external laser while $\gamma$ is the decay rate for photons in the cavity.}
\end{center}
\end{figure}

We imagine that we start with a BEC of atoms inside of an
optical cavity, as depicted in Fig. \ref{Fig1}. The atoms as well as the
molecules formed from them can be trapped inside of the cavity
using a far-off resonant optical trap similar to what has been
recently demonstrated with single atoms in a cavity \cite{boca}.
At temperatures $T\approx 0$, we can assume that all of the atoms are in the
ground state of the trapping potential with wave function
$\psi_a(r)$. Additionally, the atoms are assumed to have all been
prepared in the same hyperfine state denoted by $|a\rangle$. Pairs
of atoms in $|a\rangle$ are coupled to electronically excited
molecular states $|I_{\nu}\rangle$, where $\nu$ denotes the
vibrational state of the molecule, via a pump laser with Rabi
frequency $\Omega_l$ and frequency $\omega_l$. The pump is treated
as a large amplitude undepleted source and therefore changes in
$\Omega_l$ due to absorption or stimulated emission are
neglected.

The excited molecular states are strongly coupled to molecules in
their electronic and vibrational ground state, $|b \rangle$, via a
single cavity mode. Emission of a photon into the cavity
mode takes a molecule from an excited state to its
electronic ground state. Coupling to a single mode can be achieved by insuring that only a
single cavity mode is close to two-photon resonance for the atom-molecule Raman transition
and by positioning the atoms and molecules around an
antinode of the cavity field. The discrete mode structure of the
cavity allows one to select a particular vibrational state in the
electronic ground-state manifold of the molecules provided the cavity linewidth,
$\gamma$, is less than the vibrational level spacing, which is on
the order of $1$GHz \cite{tsai}. This would imply a cavity
Q-factor of $Q\gg 10^6$, which has already been achieved with
individual atoms trapped inside of a Fabrey-Perot
resonator\cite{boca}. The cavity field frequency is $\omega_c$ and
the vacuum Rabi frequency for the $|I_{\nu}\rangle \rightarrow
|b\rangle$ transition is $g_{cav}$. Selective emission into a
cavity mode offers a distinct advantage over using free space
spontaneous emission to populate the electronic ground state of
the molecules, since spontaneous emission would populate a large
number of vibrational levels of the ground state as well as
leading to dissociation back into the continuum
\cite{nikolov,jones}. This selectivity along with the enhanced
emission rate in a cavity due to the Purcell effect in the weak
coupling regime was proposed as a possible matter wave amplifier
for molecules \cite{search-mol-amp}.

The internal energies of states $|I_{\nu}\rangle$ and $|b\rangle$
relative to pairs of atoms in $|a\rangle$ are $\omega_{\nu}$ and
$\Delta\omega<0$, respectively. We assume that the detuning
between the excited states and the pump and cavity fields satisfy,
\[
\Delta_{\nu}=\omega_{\nu}-\omega_l\approx
(\omega_{\nu}-\Delta\omega)-\omega_c \gg
|g_{cav}|,|\Omega_l|,\gamma_{\nu}
\]
where $\gamma_{\nu}^{-1}$ is the lifetime of $|I_{\nu}\rangle$
due to spontaneous emission. Under these conditions the excited
state can be adiabatically eliminated, leading to two-photon Raman
transitions between $|a\rangle$ and $|b\rangle$ with Rabi
frequency
\[
\chi(x)\approx g_{cav}
u(x)\Omega_l^*(x)\sum_{\nu}I^*_{a,\nu}I_{b,\nu}/\Delta_{\nu}.
\]
under the assumption that the cavity mode $u(x)$ and $\Omega_l(x)$
are essentially constant over the spatial extent of the
internuclear wave functions for the molecules and colliding atoms
\cite{drummond-STIRAP}. $I_{\ell,\nu}$ are the Frank-Condon
factors for the $|\ell=a,b\rangle \rightarrow |I_{\nu}\rangle$
transitions.

At zero temperature we can use a single-mode approximation for the
molecular field as well as for the atoms. The ground state wave
function for the center of mass of the molecules is denoted by
$\psi_b(x)$. This approximation for the atomic and molecular field
should remain valid at least for times long enough to observe
dynamics lasting several Rabi periods. For longer times, one can expect that
some molecules and atoms will be created with excited state wave
function \cite{goral}. The resulting Hamiltonian for the
atom-cavity-molecule system is
\begin{equation}
\hat{H}_{atom-mol}=(\Delta\omega-(\omega_{l}-\omega_c))\hat{b}^{\dagger}\hat{b}+i\frac{\hbar
g}{2}(\hat{a}^{\dagger
2}\hat{b}\hat{e}-\hat{a}^{2}\hat{b}^\dagger\hat{e}^\dagger)
+\hbar\chi_a\hat{a}^{\dagger 2} \hat{a}^2+\hbar\chi_b
\hat{b}^{\dagger 2}\hat{b}^2
\end{equation}
where $\hat{a}$, $\hat{b}$, and $\hat{e}$ are bosonic annihilation
operators for atoms, ground state molecules, and cavity photons,
respectively. Moreover, the molecular operators and photon
operators have been written in a rotating frame,
$\hat{b}\rightarrow \hat{b}\exp[+i(\omega_c-\omega_l)t]$ and
$\hat{e}\rightarrow \hat{e}\exp[-i\omega_ct]$, to remove all time
dependence from the interaction term. The coupling constant is
given by
\[
ig=\int d^3x \psi_b^*(x)\chi(x)\psi_a(x).
\]
We can further simplify things by assuming that the two-photon
resonance condition, $\Delta\omega=\omega_l-\omega_c$, is
satisfied at all times so that the first term in $\hat{H}_{atom-mol}$ is identically zero.
The terms proportional to $\chi_a$ and $\chi_b$ represent the two
body interactions between pairs of atoms and pairs of molecules, respectively. We have not explicitly included two-body interactions
involving an atom interacting with a molecule because such a term can be written as
\[
\chi_{ab}\hat{N}_a\hat{N}_b=\frac{1}{4}\chi_{ab}\left( \hat{N}^2-\hat{N}_a^2-4\hat{N}_b^2\right)
\]
where $\hat{N}=\hat{N}_a+2\hat{N}_b$ is the total number operator and $\hat{N}_a=\hat{a}^{\dagger}\hat{a}$
and $\hat{N}_b=\hat{b}^{\dagger}\hat{b}$. Since the total number of particles is conserved, we are dealing with eigenstates
of $\hat{N}$ and the atom-molecule interaction can be absorbed into a redefinition of the atom-atom and molecule-molecule interactions.

The dynamics of the empty cavity are described by two
competing processes. The first process is cavity decay, which can
be treated using the standard Born-Markov master equation for the
density operator \cite{QuantumNoise},
\begin{equation}
\frac{d\rho}{dt}|_{damping}=\frac{\gamma}{2} \left(
2\hat{e}\rho\hat{e}^{\dagger}-\hat{e}^{\dagger}\hat{e}\rho-
\rho\hat{e}^{\dagger}\hat{e} \right)
\end{equation}
In addition to this, the cavity is pumped by some external
source. Coherent driving of the cavity by a
classical source such as a laser is described by the following interaction
Hamiltonian,
\begin{equation}
H_{pump}=i\hbar(\varepsilon\hat{e}^\dagger-\varepsilon^*\hat{e}).
\end{equation}
The complete equation of motion for the density operator
is then given by,
\begin{equation}
\frac{d\rho}{dt}=\frac{1}{i\hbar}[H_{pump}+H_{atom-mol},\rho]+\frac{\gamma}{2}
\left( 2\hat{e}\rho\hat{e}^{\dagger}-\hat{e}^{\dagger}\hat{e}\rho-
\rho\hat{e}^{\dagger}\hat{e} \right) \label{masterequation}
\end{equation}.

The quantum dynamics of the system can be analyzed by deriving an equation of
motion for the positive-P distribution,
$P(\alpha,\beta,e,\alpha^*,\beta^*,e^*,t)$, which is a
representation of the density operator in terms of bosonic
coherent states,
\begin{equation}
\rho=\int d^2\alpha d^2\alpha^* d^2\beta d^2\beta^* d^2e d^2e^*
P(\alpha,\beta,e,\alpha^*,\beta^*,e^*,t)
\frac{|\alpha,\beta,e\rangle \langle
\alpha^*,\beta^*,e^*|}{\langle
\alpha^*,\beta^*,e^*|\alpha,\beta,e\rangle}
\end{equation}
where $\hat{C}|\alpha,\beta,e\rangle=c|\alpha,\beta,e\rangle$
where $\hat{C}={\hat{a},\hat{b},\hat{e}}$ and
$c={\alpha,\beta,e}$, respectively. Note that in general ${\alpha,\beta,e}$
and ${\alpha^*,\beta^*,e^*}$ are to be understood as independent
complex variables and not simply complex conjugates.
Using Eq.(\ref{masterequation}) and standard methods
\cite{QuantumNoise}, a Fokker-Planck equation can be derived for
the positive P-distribution $P(\alpha,\beta,e,t)$.
\begin{eqnarray}\label{FokkerPlanck}
\frac{\partial P }{\partial t} &=& \Big{[}
\frac{\partial}{\partial\alpha}
(2i\chi_a\alpha^*\alpha^2-g\alpha^*\beta e) \nonumber \\
&+&\frac{\partial}{\partial\alpha^*}
(-2i\chi_a\alpha\alpha^{*2}-ge^*\beta^*\alpha)+
\frac{\partial}{\partial\beta}
(2i\chi_b\beta^*\beta^2+\frac{g}{2}\alpha^2 e^*) \nonumber \\
&+& \frac{\partial}{\partial\beta^*}
(-2i\chi_a\alpha\alpha^{*2}+\frac{g}{2}e\alpha^{*2})+ \frac{1}{2}(
\frac{\partial^2}{\partial\alpha^2}(-2i\chi_a\alpha^2+g\beta e )
\nonumber \\
&+&
\frac{\partial^2}{\partial\alpha^{*2}}(-2i\chi_a\alpha^{*2}+g\beta^*
e^* ) + \frac{\partial^2}{\partial\beta^2}(-2i\chi_b\beta^2 )
\nonumber \\
&+& \frac{\partial^2}{\partial\beta^{*2}}(2i\chi_b\beta^{*2} )
+\frac{\partial^2}{\partial\beta\partial{e}}(-g\alpha^2 )
+\frac{\partial^2}{\partial\beta^*\partial{e}^*}(-g\alpha^{*2} ) )
\Big{]}P
\end{eqnarray}
Unlike the equations of motion for single photon cavity assisted photoassociation derived in \cite{olsen1,olsen2}, there are
no third order derivatives with respect to the fields in Eq. \ref{FokkerPlanck}. Mathematically this originates from the different
positions of the cavity field annihilation and creation operators in the interaction Hamiltonian. In \cite{olsen1,olsen2}, the interaction was
of the form $\hat{a}^{2}\hat{b}^\dagger\hat{e}+h.c.$ while here it is $\hat{a}^{2}\hat{b}^\dagger\hat{e}^\dagger+h.c.$.

The most common way to solve Eq. \ref{FokkerPlanck} is to map ${\alpha,\beta,e}$ and ${\alpha^*,\beta^*,e^*}$ onto a set of Ito stochastic differential
equations \cite{QuantumNoise}. Here we are interested only in the dynamics of the mean fields, $\langle \hat{C} \rangle= \bar{c}$, which leads to deterministic c-number differential equations, which can be obtained from the Ito stochastic differential equations by setting all noise terms equal to zero. The resulting equations of motion are:
\begin{eqnarray}
\dot{\bar{\alpha}}=-2i\chi_a\bar{\alpha}^{*}\bar{\alpha}^2+g\bar{\alpha}^{*}\bar{\beta}\bar{e} \label{a-eq} \\
\dot{\bar{\beta}}=-2i\chi_b\bar{\beta}^2\bar{\beta}^{*}-\frac{g}{2}\bar{\alpha}^2\bar{e}^* \label{b-eq} \\
\dot{\bar{e}}=\varepsilon-\frac{\gamma}{2}\bar{e}-\frac{g}{2}\bar{\alpha}^2\bar{\beta}^* \label{c-eq}
\end{eqnarray}
The next section analyzes the solutions of Eqs. \ref{a-eq}-\ref{c-eq}.

\section{Stationary States and Stability}\label{THEORY}

There exists two distinct cases of stationary solutions in the system. First, there is the regime of weak driving, when the stationary value of the cavity field equals zero. The opposite case occurs when the driving is strong enough to determine the cavity amplitude regardless of the values of the matter fields.

First we treat the case of weak driving with nonzero stationary values of the atom and molecule modes, and rewrite Eqs.(\ref{a-eq}-\ref{c-eq}) in polar form using
\begin{eqnarray}
\alpha(t)=\sqrt{N-2N_b(t)}e^{i\Theta_a(t)},\\
\beta(t)=\sqrt{N_b(t)}e^{i\Theta_b(t)}.
\end{eqnarray}
Substituting these into Eqs.(\ref{a-eq}-\ref{c-eq}) we arrive at
\begin{equation}\label{SCl1}
\dot{\Delta} = 4\chi_aN-2(\chi_b+4\chi_a)N_b+g\frac{N-6N_b}{2\sqrt{N_b}}[e_I\cos(\Delta)+e_R\sin(\Delta)]
\end{equation}
\begin{equation}\label{SCl2}
\dot{N_b}=-g\sqrt{N_b}(N-2N_b)[e_R\cos(\Delta)-e_I\sin(\Delta)],
\end{equation}
\begin{equation}\label{SCl3}
\dot{e}_R=\varepsilon-\frac{\gamma}{2}e_R-\frac{g}{2}\sqrt{N_b}(N-2N_b)\cos(\Delta)
\end{equation}
\begin{equation}\label{SCl4}
\dot{e}_I=-\frac{\gamma}{2}e_I+\frac{g}{2}\sqrt{N_b}(N-2N_b)\sin(\Delta),
\end{equation}
where $e_R$ and $e_I$ are the real and imaginary parts of the cavity field mode, and where $\Delta=\Theta_b-2\Theta_b$. We note that the photo-association dynamics, as given by Eqs.(\ref{SCl1})-(\ref{SCl4}), only depends on the relative phase between atom and molecule modes, $\Delta$, and the phase dynamics of the individual modes is thus of no interest in this situation.

As a first case we investigate the possibility of a nontrivial steady state with empty cavity $e_R=e_I=0$.
From Eq.(\ref{SCl4}) we see that $\sin(\Delta)=0$ while Eq.(\ref{SCl3}) give $\cos(\Delta) > 0$, which combined show that $\Delta = 0$.
From Eq.(\ref{SCl1}), we see that in general, all four equations cannot be satisfied if $e=0$, but if the nonlinear term is neglected, a steady state is possible.
Rewriting Eq.(\ref{SCl3}) in terms of the molecule fraction,
\[
 z^2=2N_b/N,
 \]
we find the condition for stationarity
\begin{equation}\label{Pol3}
P_3(z)\equiv z^3-z+C=0,
\end{equation}
where the dimensionless constant $C$ is given by
\begin{equation}\label{C_def}
C=\frac{\varepsilon}{g}\left(\frac{N}{2}\right)^{-3/2}.
\end{equation}
and represents the ratio of the rates for the two competing processes that populate the cavity with photons- pumping and emission of a photon during the atom to molecule Raman transition. In terms of $C$, weak driving therefore corresponds to $C\ll 1$ while strong driving is $C\gg 1$.

Equation Eq.(\ref{Pol3}) has two roots in the interval $0\le z\le 1$ if
\begin{equation}\label{C_condition}
C \le C_{crit} = \frac{2}{3\sqrt{3}},
\end{equation}
where the upper limit $C_{crit}$ denotes the value where $\min P_3(z)=0$, \textit{i.e.} when the two roots coincide, which occurs for $z=1/\sqrt{3}$.
In order to examine the stability of the stationary solutions, we linearize Eqs.(\ref{SCl1})-(\ref{SCl4}) for  small perturbations around the stationary values
\begin{eqnarray}
\Delta(t)\approx\Delta_0+\delta{\Delta}(t),\\
N_b(t)\approx N_0+\delta{N_b}(t),\\
e_R(t)\approx e_{R,_0}+\delta{e_R}(t),\\
e_I(t)\approx\ e_{I,0}+\delta{e_I}(t),
\end{eqnarray}
where $\Delta_0=e_{R,0}=e_{I,0}=0$ and $N_0$ is one of the two roots of Eq. (\ref{Pol3}).
The linearized equations of motion are given by
\begin{equation}\label{EqSyst}
\left [ \begin{array}{ c }
\dot{\delta\Delta}\\
\dot{\delta N_b}\\
\dot{\delta e_R}\\
\dot{\delta e_I}
\end{array}\right] =
\left [\begin{array}{ c c c c}
0 & 0 & 0 & -2A \\
0 & 0 & -\frac{\varepsilon}{2} & 0 \\
0 & A & -\frac{\gamma}{2} & 0 \\
\varepsilon & 0 & 0 & -\frac{\gamma}{2}
\end{array}\right ]
\left [ \begin{array}{ c }
\delta\Delta\\
\delta N_b\\
\delta e_R\\
\delta e_I
\end{array}\right],
\end{equation}
where the nonlinear interactions have been dropped and
\begin{equation}
A=-\frac{g}{2}\frac{N-6N_0}{2\sqrt{N_0}}.
\end{equation}
The eigenvalues of the Jacobian in Eq.(\ref{EqSyst}) are given by
\begin{equation}
\lambda_{1,2}=-\frac{\gamma}{4} \left [ 1\pm\sqrt{1-32\varepsilon A/\gamma^2} \right ],
\end{equation}
and
\begin{equation}
\lambda_{3,4}=-\frac{\gamma}{4}\left [ 1\pm\sqrt{1-8\varepsilon A/\gamma^2} \right ],
\end{equation}
respectively.
Stability requires that the Jacobian has all eigenvaules with negative real parts, a condition which implies
$N_0>N/6$, or equivalently, $z^2 > 1/3$. From this analysis we can conclude that for $C<C_{crit}$, one root, $z_+>1/\sqrt{3}$, is stable while the other root, $z_{-}<1/\sqrt{3}$, is unstable. At $C=C_{crit}$, these two roots merge and stability is completely lost as the system passes through a saddle node bifurcation.

\begin{figure}[ht]
\begin{center}
\includegraphics[width=8.5cm]{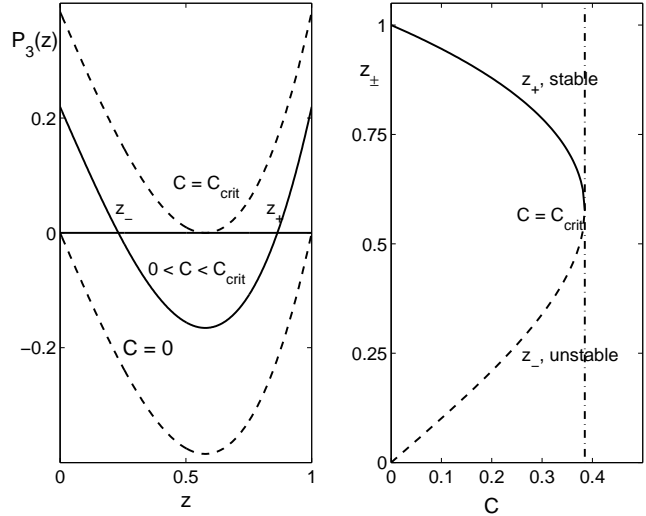} \caption{\label{fig2}Stability plot for the regime with empty cavity, $e=0$ and $\chi_a=\chi_b=0$. The left panel shows $P_3(z)$, whose roots give the stationary molecule fraction. The right pane shows the real roots of $P_3(z)$, where the upper branch, $z_{+}$, is stable for $C \leq C_{crit}$ while the lower branch, $z_{-}$, is unstable.}
\end{center}
\end{figure}

\begin{figure}[ht]
\begin{center}
\includegraphics[width=8.5cm]{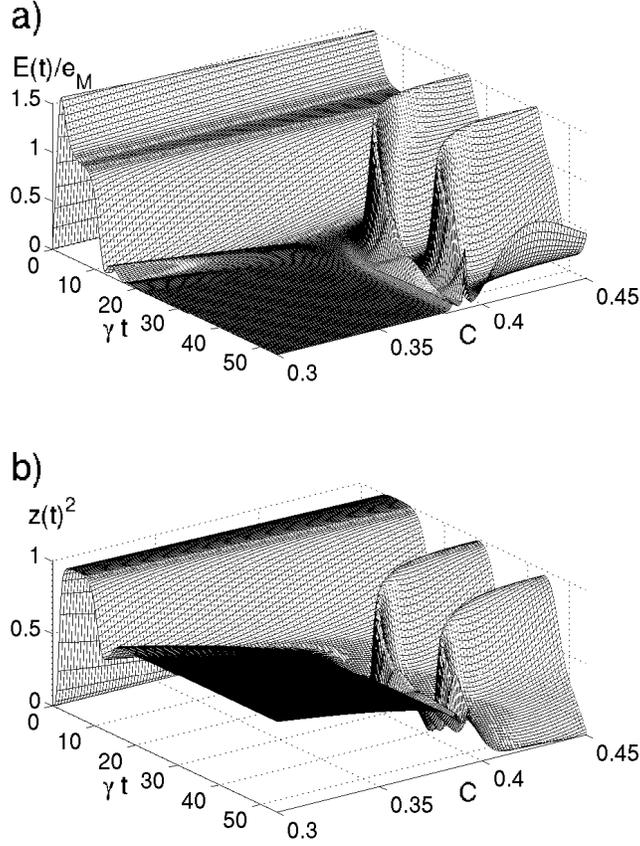} \caption{\label{fig3}Numerical solution of Eqs.(\ref{SCl1}-\ref{SCl4}). In a) the cavity field amplitude $E(t)$ is plotted as a function of time for different values of the parameter $C$. the amplitude is scaled in units of $e_M = 2\varepsilon/\gamma$. In b) the molecule fraction $z^2(t)$ is shown as function of time for different values of the parameter $C$. At $C=C_{crit}\approx 0.38$, the steady state disappears and large amplitude coherent oscillations between the atomic and molecular fields begin. This is accompanied by a nonzero cavity field. In these simulations $\chi_a=\chi_b=10^{-6}$ in units of $\gamma$, small enough to be negligible in the regime $C<C_{Crit}$.}
\end{center}
\end{figure}

As one can see from Eqs. (\ref{SCl3})-(\ref{SCl4}), the factor $g\left(N/2\right)^{3/2}$ corresponds to the maximal atom-molecule polarization that absorbs energy from the cavity mode (assuming $N_b\approx N$). In order to have $e=0$ stationary, this coherence term in Eq.(\ref{SCl3}) has to balance the driving term given by $\varepsilon$. The constant $C$ is a measure then of the ratio between these two terms, and when $C>C_{Crit}$, the driving is too strong to be compensated for by the absorption, and the steady state can no longer be maintained. Fig. (\ref{fig3}) shows the cavity amplitude and molecule fraction as a function of time for different values of $C$. As one can see for $C>C_{crit}$, there are large amplitude nonlinear Rabi oscillations between the atomic and molecular fields.

We focus next on the case of strong driving, \textit{i.e.} $C \gg C_{crit}$, which leads to a nonempty cavity. Representing the cavity field using a polar representation,
\begin{equation}
e(t)=E(t)e^{i\Theta_e(t)},
\end{equation}
leads to the following dynamical equations
\begin{equation}\label{ENZ1}
\dot{\Delta} = 4\chi_aN-2(\chi_b+4\chi_a)N_b+g\frac{N-6N_b}{2\sqrt{N_b}}E\sin(\Delta+\Theta_e),
\end{equation}
\begin{equation}\label{ENZ2}
\dot{N_b}=-g\sqrt{N_b}(N-2N_b)E\cos(\Delta+\Theta_e),
\end{equation}
\begin{equation}\label{ENZ3}
\dot{\Theta}_e=-\frac{\varepsilon}{E}\sin(\Theta_e)+\frac{g}{2E}\sqrt{N_b}[N-2N_b]\sin(\Delta+\Theta_e),
\end{equation}
and
\begin{equation}\label{ENZ4}
\dot{E}=\varepsilon\cos(\Theta_e)-\frac{\gamma}{2} E-\frac{g}{2}\sqrt{N_b}[N-2N_b]\cos(\Delta+\Theta_e),
\end{equation}
for the amplitudes and phases. Again we find the stationary values corresponding to $\dot{\Delta}=\dot{N}_b=\dot{E}=\dot{\Theta}_{e}=0$ in terms of the physical parameters. Here, that fact that $E > 0 $ and $0 < N_b < N/2$ in the stationary state is used to assist in finding nontrivial solutions. First, for $\dot{N_b} = 0$, we get from Eq.(\ref{ENZ2}) that $\cos(\Delta+\Theta_e) = 0$, which gives $\sin(\Delta+\Theta_e) = \sigma =  \pm1$. Using this in Eq.(\ref{ENZ3}) gives, together with Eq.(\ref{C_def})
\begin{equation}
\sin(\Theta_e) = \sigma\frac{g}{2\varepsilon}\left(\frac{N}{2}\right)^{3/2}z\left(1-z^2\right) = \sigma\frac{1}{2C}z\left(1-z^2\right).
\end{equation}
In the regime of strong pumping $C \gg 1$ we have $|\sin(\Theta_e)| < \frac{2}{C} \ll 1$, which in turn implies $\cos(\Theta_e)\approx 1$.
These results when inserted in Eq.(\ref{ENZ4}), yield the cavity amplitude
\begin{equation}
E=\frac{2\varepsilon}{\gamma} = e_{M}, \label{ENZE}
\end{equation}
which is the maximal sustainable field amplitude and is what one would expect for a cavity with no atoms or molecules in the cavity.

\begin{figure}[ht]
\begin{center}
\includegraphics[width=8.5cm]{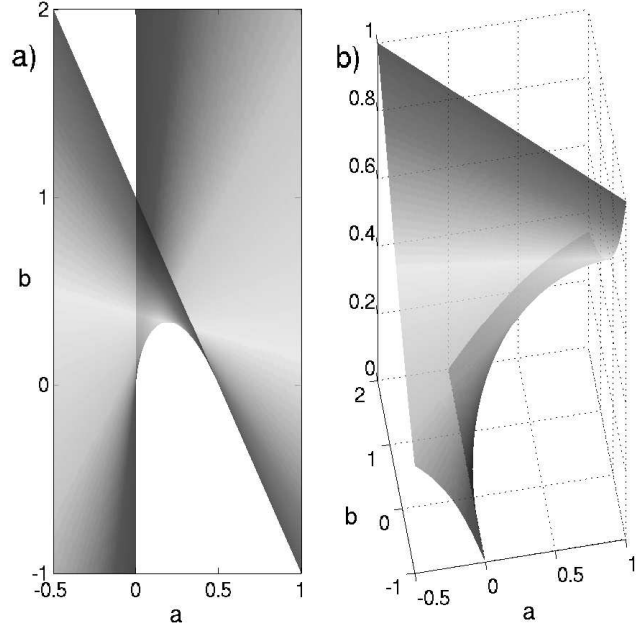} \caption{\label{fig4} The real roots of Eq. \ref{P3ab}, which gives the surface of stationary molecule fraction values  as function of $a$ and $b$ for the regime with non-empty cavity, $E_0 >0$. Note that the left pane, (a), and the right pane, (b), show the same surface but for from different viewing perspectives.}
\end{center}
\end{figure}

Using Eq.(\ref{ENZE}) in Eq.(\ref{ENZ1}), we finally arrive at a cubic polynomial for the molecule number
\begin{equation}\label{P3ab}
z^3-3az^2-bz+a=0;
\end{equation}
where
\begin{eqnarray}
a&=&\sigma C\frac{g^2N}{2\gamma}\frac{1}{4\chi_a-\chi_b}, \\
b&=&\frac{1}{1-\frac{\chi_a}{4\chi_b}}.
\end{eqnarray}
The stationary molecule fractions thus only depend on the two parameters $a$ and $b$ for strong pumping. The solution manifold for Eq.(\ref{P3ab}), \textit{i.e.} the surface of possible values of  the molecule fraction $ 0\leq z(a,b) \leq 1$ is shown in Fig. \ref{fig4}. Here we see that there is a small region around $0\leq a \leq 0.5$, $0\leq b \leq 1$ where Eq.(\ref{P3ab}) has two roots, which indicates the possible occurrence of bistability.

Next, we turn to the question of stability of the stationary solutions given by Eq.(\ref{P3ab}). Linearizing Eqs. (\ref{ENZ1})-(\ref{ENZ4}) for small perturbations around the stationary points
\begin{eqnarray}
\Delta(t)\approx\Delta_0+\delta{\Delta}(t),\\
N_b(t)\approx N_0+\delta{N_b}(t),\\
E(t)\approx E_0+\delta{E}(t),\\
\Theta_e(t)\approx\ \Theta_{e,0}+\delta\Theta_e(t),
\end{eqnarray}
we get the Jacobian
\begin{equation}\label{Jac2}
J=
\left [\begin{array}{ c c c c}
0 & B_1 & 0 & B_3 \\
B_2 & 0 & B_2 & 0 \\
0 & -\frac{\gamma B_3}{4\varepsilon} & -\frac{\gamma}{2} & 0 \\
\frac{B_2}{2} & 0 & 0 & -\frac{\gamma}{2}
\end{array}\right ]
\end{equation}
where the matrix elements are given by
\begin{eqnarray}\label{B1}
B_1&=&\chi-\sigma\frac{2\varepsilon g}{\gamma}\left(\frac{N}{2}\right)^{-\frac{1}{2}}f_1(z), \\
\label{B2}
B_2&=&\sigma2g\left(\frac{N}{2}\right)^{\frac{3}{2}}f_2(z), \\
\label{B3}
B_3&=&\sigma\frac{g}{\sqrt{2}}\left(\frac{N}{2}\right)^{\frac{1}{2}}f_3(z),
\end{eqnarray}
and where we introduced the functions $f_1(z)=(1+3z^2)/z^3$, $f_2(z)=z(1-z^2)$, and $f_3(z)=(1-3z^2)/z$, as well as the interaction constant $\chi = 2(4\chi_a-\chi_b)$.
The resulting secular equation for the eigenvalues is
\begin{equation}\label{SecEq}
\lambda^4+\lambda^3+\left[\frac{1}{4}+(e_M-1)A-B\right]\lambda^2+\left[\frac{e_M-1}{2}A-B\right]\lambda-\left[B+\frac{A^2}{e_M}\right]=0.
\end{equation}
where
\begin{equation}\label{A}
A=\frac{B_2B_3}{\gamma^2}
\end{equation}
and
\begin{equation}\label{B}
B=\frac{B_1B_2}{\gamma^2}.
\end{equation}
Using the Routh-Hurwitz stability criterion on Eq.(\ref{SecEq}), we find that necessary conditions  for the eigenvalues of the Jacobian to have non-positive real parts are
\begin{equation}\label{ACond1}
A\geq -\frac{1}{2(e_M-1)},
\end{equation}
and
\begin{equation}\label{ACond2}
\frac{A^2}{e_M}\geq -B,
\end{equation}
and
\begin{equation}\label{ACond3}
\left[\frac{1}{4}(e_M-1)^2-\frac{1}{e_M}\right]A^2-\frac{1}{4}(e_M-1)\left[\frac{1}{4}+B\right] A-\frac{3}{4}B\leq0
\end{equation}
The Routh-Hurwitz stability criterion is in general only necessary, and not always sufficient \cite{mabuchi}. Here we find, however, that the two conditions (\ref{ACond2}) and (\ref{ACond3}) are sufficient as a numerical check confirms that the roots of Eq.(\ref{SecEq}) indeed have non-positive real parts when (\ref{ACond2}) and (\ref{ACond3}) are fulfilled. In addition, we find that Eq(\ref{ACond2}) corresponds to loss of stability through a saddle-node bifurcation, whereas Eq.(\ref{ACond3}) corresponds to a Hopf-bifurcation as two roots of Eq.(\ref{SecEq}) cross into the right half of the complex plane.

\begin{figure}[ht]
\begin{center}
\includegraphics[width=8.5cm]{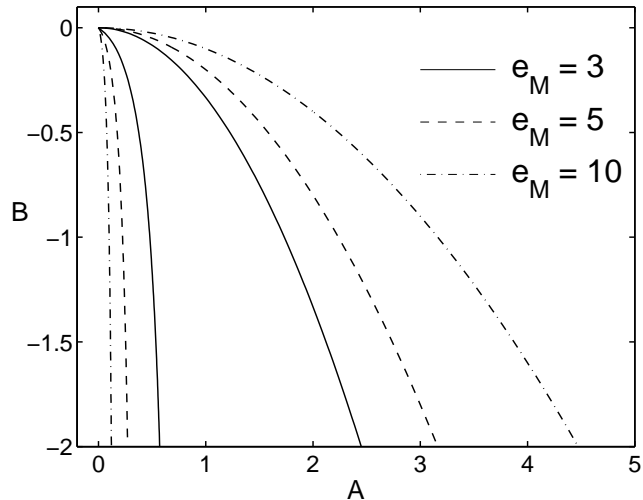}\caption{\label{BLines} Region of stability in the $A$-$B$ plane for different values of $e_M$. The stability diagram was obtained from the Routh-Hurwitz criterion and explicitly checked by direct numerical solution of the roots of Eq. \ref{SecEq}. The region enclosed by the left and right borders represents the stable region where all roots have negative real parts. On the left boundary of this region there is a Hopf bifurcation and on the right boundary, a saddle node bifurcation.}
\end{center}
\end{figure}

We first note that in general Eq.(\ref{ACond1}) is trivially fulfilled when the two inequalities (\ref{ACond2})-(\ref{ACond3}) are fulfilled.
Rewriting Eq.(\ref{ACond2}) in terms of the physical parameters we get
\begin{equation}\label{messy}
-2\sigma\frac{\chi\gamma}{g\varepsilon}\left(\frac{N}{2}\right)^\frac{1}{2}
\leq
\sqrt{2}f_1(z)-
\frac{f_2f_3^2}{4C^2\gamma^2}.
\end{equation}
For the case $\gamma C\ll 1$, and $\sigma = -1$ combined with the assumption that the functions $f_i$ are limited, we have
\begin{equation}
\frac{\chi\gamma}{g\varepsilon}\left(\frac{N}{2}\right)^\frac{1}{2} \approx 1
\end{equation}
as an approximation for the stability boundary given by Eq.(\ref{ACond1}). Fig. 5 shows the region of stability in the A-B plane.
\begin{figure}[ht]
\begin{center}
\includegraphics[width=8.5cm]{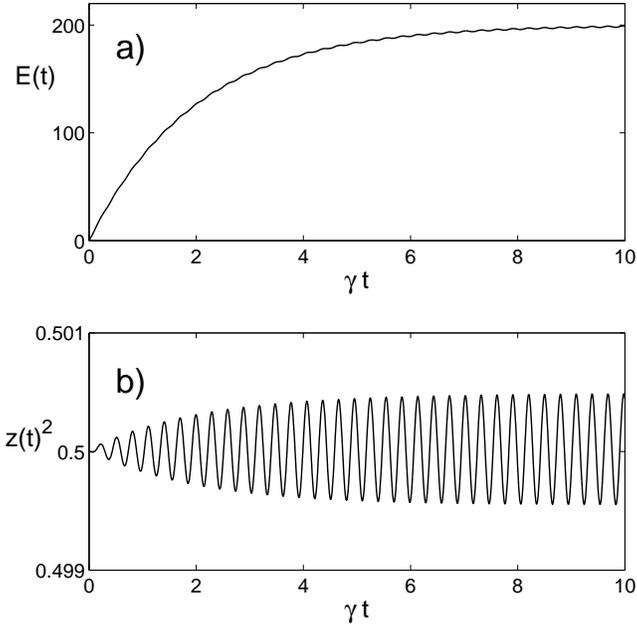} \caption{\label{Upper}Plot of time-dependence of the cavity field amplitude and molecule fraction for initial values, $E=0$ and $z=0.5$. $z=0.5$ was chosen to lie in the vicinity of the upper fold on the manifold of stationary solutions shown in Fig 4. $E(t)$ is plotted in units of $\gamma$ and $e_M=200$.}
\end{center}
\end{figure}

\begin{figure}[ht]
\begin{center}
\includegraphics[width=8.5cm]{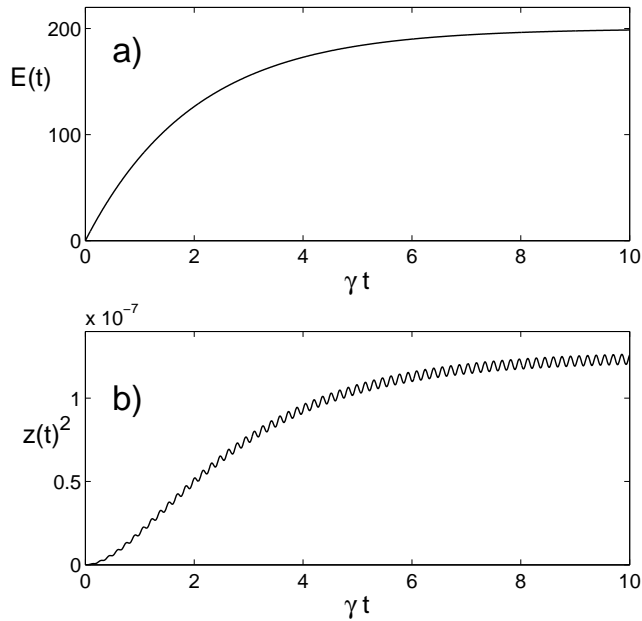} \caption{\label{Lower}Same as Fig. 6 except that the initial value for the molecule fraction was chose to be $z=0$, which lies close to the lower fold on the manifold of stationary solutions shown in Fig 4.}
\end{center}
\end{figure}

These results indicate that for specific choices of the parameters, both stationary states will be stable.
As an example of the dynamics of the mean fields near the stationary states, we solved Eq.(\ref{P3ab}) for $a\approx 3.3 \times 10^{-4}$, b=$1/2$ giving $z\approx 0.5$ and $z \approx 10^{-7}$. The values $z=0.5$ and $z=0$ were used as initial conditions in Eqs.(\ref{ENZ1})-(\ref{ENZ4}). The results are shown in Figs. \ref{Upper} and \ref{Lower} for the cavity field and molecule fraction as functions of time. In Figs. \ref{Upper} (a) and \ref{Lower} (a), one sees that the cavity field amplitude approach $e_M$ exponentially for large times. The molecule fractions are shown in Figs. \ref{Upper} (b) and \ref{Lower} (b). For both cases, the solutions exhibit small amplitude oscillations around the stationary values, something characteristic of the presence of Hopf bifurcations.

\section{Summary and Conclusions}

We have investigated the semiclassical dynamics of cavity assisted photo-association of molecules. Two different cases were described in some detail, namely weak and strong pumping. For weak pumping and zero two-body interactions, the steady state cavity field is zero. From Eqs. (\ref{SCl3})-(\ref{SCl4}) we see that for
a stationary state with $e=0$, the pumping has to be exactly balanced by the atom-molecule coherence, which causes absorption.
We interpret this regime as that where the number of photons in the cavity is smaller than or
equal to the maximum number of molecules. All of the photons injected into the cavity by the driving are used to convert a fraction of the atoms
to molecules thereby leaving the cavity empty. When the pumping reaches a critical value, the system undergoes a saddle-node bifurcation and stability is lost. At this point the number of photons is greater than the maximum number of molecules and the system is able to undergo full amplitude nonlinear Rabi oscillations without
fully depleting the cavity field. For strong pumping and non-zero two-body interactions, we find that there exists bistability in the molecule number in the prescence of a nonzero steady state cavity field. The bistable steady state molecule fractions are determined by the relative values of the atom-atom and molecule-molecule two-body interactions. The boundaries of the bistable region are marked by a saddle node bifurcation and a Hopf bifurcation.

In a future publication we intend to go beyond the mean-field dynamics to explore the entanglement of cavity photons with molecular dimers. The form of $\hat{H}_{atom-mol} \propto i\hbar g(\hat{a}^{\dagger2}\hat{b}\hat{e}-\hat{a}^{2}\hat{b}^\dagger\hat{e}^\dagger)/2$ is of the same form as the four-wave mixing terms that give rise to optical phase conjugation and two-photon squeezed states \cite{bondurant}.


\begin{thebibliography}{99}

\bibitem{mol-1} E. A. Donley, N. R. Claussen, S. T. Thompson, and C. E.
Wieman, Nature (London) {\bf 417}, 529 (2002).

\bibitem{durr} S. Durr, T. Volz, A. Marte, and G. Rempe, Phys. Rev. Lett. {\bf 92},
020406 (2004).

\bibitem{xu-2003} K. Xu, T. Mukaiyama, J. R. Abo-Shaeer, J. K. Chin, D. E.
Miller, and W. Ketterle, Phys. Rev. Lett. {\bf 91}, 210402 (2003).

\bibitem{herbig} Jens Herbig, Tobias Kraemer, Michael Mark, Tino Weber, Cheng
Chin, Hanns-Christoph N{\"a}gerl, Rudolf Grimm, Science {\bf 301},
1510 (2003).

\bibitem{regal-2003} C. A. Regal, C. Ticknor, J. L. Bohn, and D. S. Jin, Nature
(London) 424, 47 (2003)

\bibitem{strecker-2003} K. E. Strecker, G. B. Partridge, and R. G. Hulet, Phys. Rev.
Lett. 91, 080406 (2003).

\bibitem{jochim-2003} S. Jochim et al., Phys. Rev. Lett. 91, 240402 (2003).

\bibitem{cubizolles-2003} J. Cubizolles et al., Phys. Rev. Lett. 91, 240401 (2003).

\bibitem{mol-BEC-K} M. Greiner, C. A. Regal, and D. S. Jin, Nature (London) 426,
537 (2003).

\bibitem{mol-BEC-Li} M. W. Zwierlein, C. A. Stan, C. H. Schunck, S. M. F. Raupach,
S. Gupta, Z. Hadzibabic, W. Ketterle, Phys. Rev. Lett. 91, 250401
(2003);  S. Jochim, M. Bartenstein, A. Altmeyer, G. Hendl, S.
Riedl, C. Chin, J. Hecker Denschlag, Science 302, 2101 (2003).

\bibitem{wynar} R. Wynar R.S. Freeland, D.J. Han, C. Ryu, D.J. Heinzen, Science {\bf 287}, 1016 (2000);

\bibitem{julienne1} P. S. Julienne, K. Burnett, Y. B. Band, and W. C. Stwalley,
Phys. Rev. A {\bf 58}, R797 (1998); D. Heinzen, R. Wynar, P.
Drummond, and K. Kheruntsyan, Phys. Rev. Lett. {\bf 84}, 5029
(2000).

\bibitem{rom} Tim Rom, Thorsten Best, Mandel, Artur Widera, Markus Greiner, Theodor W. H{\"a}nsch, and Immanuel
Bloch, Phys. Rev. Lett. {\bf 93}, 073002 (2004).

\bibitem{winkler} K. Winkler, G. Thalhammer, M. Theis, H. Ritsch, R. Grimm, and J. Hecker
Denschlag, Phys. Rev. Lett. 95, 063202 (2005).

\bibitem{tsai} C. C. Tsai R. S. Freeland, J. M. Vogels, H. M. J. M. Boesten, B. J. Verhaar, and D. J. Heinzen, Phys. Rev. Lett. {\bf 79}, 1245
(1997).

\bibitem{jones} Kevin M. Jones, Eite Tiesinga, Paul D. Lett, and Paul S.
Julienne, Rev. Mod. Phys. {\bf 78}, 483 (2006).

\bibitem{yurovsky} V. A. Yurovsky, A. Ben-Reuven, P. S. Julienne,
and C. J. Williams, Phys. Rev. A {\bf 60}, R765 (1999); V. A.
Yurovsky and A. Ben-Reuven, Phys. Rev. A {\bf 72}, 053618 (2005).

\bibitem{mukaiyama} T. Mukaiyama, J. R. Abo-Shaeer, K. Xu, J. K.
Chin, and W. Ketterle, Phys. Rev. Lett. {\bf 92}, 180402 (2004).

\bibitem{molecule-optics} J. R. Abo-Shaeer, D. E. Miller, J. K. Chin, K. Xu,
T. Mukaiyama, and W. Ketterle, Phys. Rev. Lett. 94, 040405 (2005).

\bibitem{moore} M. G. Moore, O. Zobay, and P. Meystre
Phys. Rev. A {\bf 60}, 1491-1506 (1999); M. G. Moore and P.
Meystre Phys. Rev. A {\bf 59}, R1754-R1757 (1999).

\bibitem{olsen1} M. K. Olsen, J. J. Hope, and L. I. Plimak, Phys. Rev. A 64,
013601 (2001).

\bibitem{olsen2} M. K. Olsen, L. I. Plimak, and M. J. Collett, Phys. Rev. A 64,
063601 (2001).

\bibitem{QuantumNoise} C. W. Gardiner and P. Zoller, "Quantum Noise: A Handbook of Markovian
and Non-Markovian Quantum Stochastic Methods with Applications to
Quantum Optics" (Springer-Verlag, Berlin, 2004).

\bibitem{boca} A. Boca, R. Miller, K. M. Birnbaum, A. D. Boozer, J. McKeever,
and H. J. Kimble, Phys. Rev. Lett. {\bf 93}, 233603 (2004).

\bibitem{nikolov} A. N. Nikolov, E. E. Eyler, X. T. Wang, J. Li, H. Wang, W. C. Stwalley, and P. L. Gould, Phys. Rev. Lett. {\bf 82}, 703 (1999).

\bibitem{search-mol-amp} C. P. Search and Pierre Meystre Phys. Rev. Lett. {\bf 93}, 140405
(2004)

\bibitem{drummond-STIRAP} P. D. Drummond and K. V. Kheruntsyan, D. J. Heinzen and R. H.
Wynar, Phys. Rev. A {\bf 65}, 063619 (2002).

\bibitem{goral} Krzysztof Goral, Mariusz Gajda, and Kazimierz
Rzazewski, Phys. Rev. Lett. {\bf 86}, 1397 (2001).

\bibitem{mabuchi} M. A. Armen and H. Mabuchi, Phys. Rev. A {\bf 73}, 063801 (2006).

\bibitem{bondurant} Roy S. Bondurant, Prem Kumar, Jeffrey H. Shapiro, and Mari Maeda, Phys. Rev. A {\bf 30}, 343 (1984); Horace P. Yuen and Jeffrey H. Shapiro, Opt. Lett. {\bf 4}, 334 (1979).

\end{thebibliography}
\end{document}